\documentstyle[graphicx,float]{article}

\begin{document}
\title{Joint effects of entanglement and symmetrization: physical properties and
exclusion}
\date{}
\author{Pedro Sancho \\ Centro de L\'aseres Pulsados CLPU \\ Parque
Cient\'{\i}fico, 37085 Villamayor, Salamanca, Spain }
\maketitle
\begin{abstract}
Entanglement and symmetrization lead to non-separable states that
can modify physical properties. Using the example of atomic
absorption we  compare both types of effects when they are relevant
at once. The presence of multi-particle superpositions largely
alters the absorption rates of identical atoms, even inhibiting the
dependence on overlapping for fermions. We also identify a set of
non-standard excluded states related to multi-fermion superposition
that naturally emerge in this context. We propose an arrangement
based on the dissociation of molecules to test these ideas.
\end{abstract}


\section{Introduction}

There are two types of non-separability in quantum theory,
these associated with entangled and symmetrized states. Both
types have been extensively studied. However, the interplay between
them is yet an open subject. There is a vivid debate about their
relationship, mainly centered in two points: the definition of
entanglement in systems of identical particles, and the possibility
of extracting useful entanglement from symmetrized states.

With respect to the first point, it is well-known that there is
no general agreement on the definition of adequate measures of
entanglement for identical particles, in spite of the fact that
several proposals have been presented in the literature. These
proposals are based on so different ideas as the use of analogues of
the Schmidt decomposition \cite{cir}, the determination of
properties that can be ascribed to the system \cite{Gh1,fri},
the exploration of the mathematical structure of the observables
\cite{Ben}, and the introduction of states without labels for
identical particles \cite{sr}. As a consequence of this lack of
consensus, it is even discussed if some states of indistinguishable
particles are entangled or not.

In the same vein there is some controversy about the second point.
One can ask if the symmetrization of identical particles alone can
be used as a resource in physical tasks. Several authors have
proposed schemes that convert the pure symmetrization into practical
entanglement, answering in the positive this question
\cite{prb,ple,LF}. These results suggest that entanglement and
symmetrization are two manifestations of a more general phenomenon,
non-separability. However, it would be desirable a deeper and more
quantitative understanding of the relation between them when both
forms of non-factorizability are simultaneously present.

We present in this paper an analysis in this line, which is based on
a simple example, atomic absorption. The associated physical property, the
light absorption rate, is dependent on the non-separability of
atomic systems. This dependence has been studied in \cite{yap} for
symmetrized states and in \cite{yej} for entangled ones. They are
only two examples of the general dependence of light-matter
interactions on non-factorizability \cite{fic,fra,dow}. Using this
dependence we can compare situations where symmetrization acts alone with others
where simultaneously we have multi-particle superpositions
associated with entanglement. The absorption rates of identical
atoms are drastically modified, both qualitatively and
quantitatively, by the multi-particle superpositions. They can lead
to large changes of the intensity of the rates, to reverse the
analytical form of the curves or to the inhibition of the dependence
on the overlap degree for fermions.

The joint effects of entanglement and symmetrization are by no means
restricted to modifications of some physical properties. We show the
existence of a set of non-standard excluded states. They do not obey
the standard Pauli's exclusion principle \cite{pau}. Instead, they
are related to the presence of multi-fermion superpositions. In
\cite{yar} it was suggested, from a more formal perspective, that in
the case of entangled systems of identical fermions there can be
excluded states beyond the scope of Pauli's exclusion principle. Our
example shows that these hypothetical states could be present in
realistic systems.

It is possible, in principle, to test the above ideas. We propose an
experimental arrangement based on light absorption by atoms prepared in
the photodissociation of molecules. Photodissociation of molecules
composed of identical atoms has been experimentally used to study the role of
entanglement in spontaneous emission \cite{jap,bel}. Later, it was
demonstrated that the processes of disentanglement \cite{ebe} and
symmetrization must be taken into account to give a complete
description of the experiments \cite{ypr}. That type of experiment
can be easily adapted to test the simultaneous effects of symmetrization and
multi-particle superposition.

\section{The system}

First of all we describe the system considered in this paper. It
consists of pairs of identical atoms, either bosons or fermions,
prepared in the initial state
\begin{eqnarray}
|\Phi >=N_0[a(|\psi>_1|\phi>_2 \pm |\phi>_1|\psi>_2)+ \nonumber \\
b(|\varphi>_1|\chi>_2 \pm |\chi>_1|\varphi>_2)]|g>_1|g>_2
\label{eq:uno}
\end{eqnarray}
where $a$ and $b$ are two complex coefficients. The center of mass
(CM) one-particle states of the atoms are $\psi$, $\phi$, $\varphi$
and $\chi$. They are normalized, $<\psi |\psi >=1, \cdots$. On the
other hand, in general, they are not orthogonal: $<\psi|\phi> \neq
0, \cdots $. The internal or electronic states of the atoms are $g$
and $e$ that refer to the ground and excited states. In the double
sign $\pm$, the upper one holds for bosons and the lower one for
fermions. The normalization coefficient is denoted by $N_0$ and
given by
\begin{eqnarray}
N_0=(2(|a|^2+|b|^2) \pm 2|a|^2|<\psi |\phi>|^2 \pm 2|b|^2|<\psi |\phi>|^2 +
\nonumber \\ 4Re(a^*b<\psi |\varphi><\phi |\chi>) \pm 4Re(a^*b<\psi |\chi> <\phi
|\varphi>) )^{-1/2}
\end{eqnarray}

This state corresponds to the (anti)symmetrization of the
unnormalized entangled state $|\Psi>=[a|\psi>_1|\phi>_2 +
b|\varphi>_1|\chi>_2]|g>_1|g>_2$. When the state describing two identical
particles is the (anti)symmetrization of a superposition of
two-particle states we have symmetrization and entanglement effects
at once. In effect, using for instance the approach in
\cite{Gh1}, we have that the (anti)symmetrization of  $\Psi$,
given by $|\Phi>_{12} =|\Psi>_{12} \pm |\Psi>_{21}$, is in general entangled
when the two coefficients $a$ and $b$ are not null. On the other
hand, when the CM-states are not orthogonal we have overlap between
the atoms and consequently exchange effects.

After the preparation, the two atoms interact with light. The light
beam contains the absorption frequency of the atoms. The intensity
of the beam is low, we assume that the probability of more than one
absorption is negligible. This way we simplify the calculations,
avoiding two-photon  absorption or other non-linear processes.

Next, we discuss a viable realization of this arrangement. We are
looking for a scheme where the symmetrization and entanglement
effects are simultaneously relevant. In addition, we must be able to
compare the single absorption probability for various overlap
degrees between the identical atoms. Our proposal is based on the
photodissociation of molecules. This process has already been used
in two experiments \cite{jap,bel} to study entanglement in
spontaneous emission. A detailed analysis of the experiments shows
that symmetrization and entanglement (plus disentanglement
\cite{ebe}) are simultaneously necessary to explain the results
\cite{ypr}. Thus, this scheme fulfills our first demand.

\begin{figure}[H]
\center
\includegraphics[width=12cm,height=7cm]{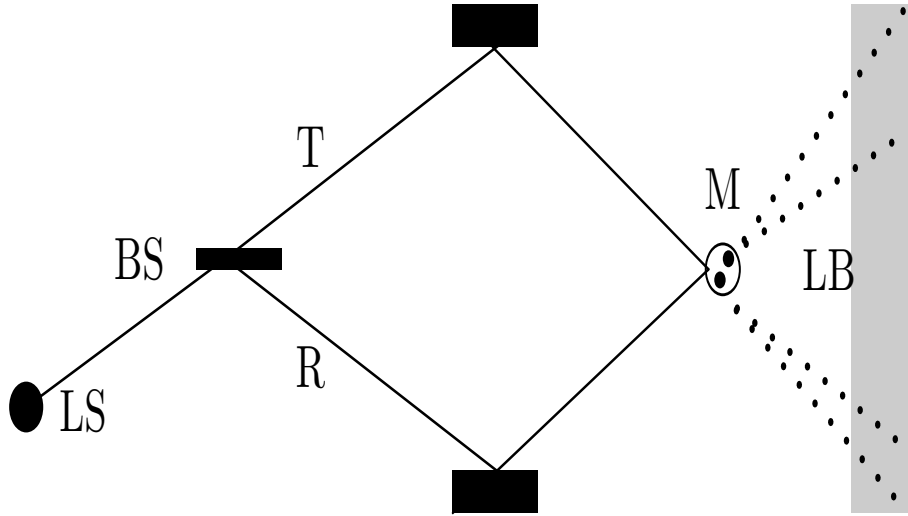}
\caption{Schematic representation of the arrangement. LS denotes a
light source, BS the beam splitter, T and S the transmitted and
reflected light rays, M the molecule composed of two identical
atoms, and LB the light beam interacting with the atoms released in
the photodissociation.}
\end{figure}

In these experiments identical atoms in excited states were prepared
by molecular photodissociation. In our case, we want to have pairs
of identical atoms, but in their ground states. In our proposal a
source produces a light beam tuned to the photodissociation
frequency of the molecule. The beam interacts with a beam splitter
leading to the superposition $|\gamma > \rightarrow a|\gamma _T> +
b|\gamma _R>$, provided that the transmission and reflection
coefficients are $a$ and $b$. Using glasses or other optical
elements the beams are directed towards the molecule. We can choose
different incidence angles for both beams (although arriving
simultaneously). This way, we can have different movement directions
and overlap degrees for the atoms. Because of the superposition of
the incident photon the two atoms after the dissociation will be in
a two-atom superposition of the type described by Eq.
(\ref{eq:uno}). In any photodissociation process the atoms show
strong quantum momentum correlations, reminiscent of the classical
law of momentum conservation.

Next, we make interact the atoms with another light beam tuned to the
excitation frequency of the atoms. If the delay between the
dissociation and the light-matter interaction is short enough the
overlap between the atoms will not be negligible. Varying the delay
we can study the variation of the absorption rate with the overlap
degree. The number of absorptions can be measured detecting the
subsequent spontaneous emissions. The scheme demands an extreme
temporal precision in the control of the delays.

\section{Absorption probabilities}

In this section we evaluate the absorption rates of the atoms after
the interaction with the light. The absorption event can happen in
four different ways: to be absorbed by an atom in the CM state
$\psi$, in the $\phi$, in the $\varphi $, or in the $\chi$. After
the absorption the CM-state changes from $\psi$, $\phi$, $\varphi $,
$\chi$ to $\psi^*$, $\phi^*$, $\varphi ^* $, $\chi ^*$ because of
the recoil. As the final state must be symmetrized we have four
different (unnormalized) states representing the absorption:
\begin{equation}
|I>=|\psi ^*>_1|e>_1 |\phi >_2|g>_2 \pm |\phi >_1|g>_1 |\psi ^* >_2|e>_2
\end{equation}
\begin{equation}
|II>=|\psi >_1|g>_1 |\phi ^*>_2|e>_2 \pm |\phi^* >_1|e>_1 |\psi >_2|g>_2
\end{equation}
\begin{equation}
|III>=|\varphi ^*>_1|e>_1 |\chi >_2|g>_2 \pm |\chi >_1|g>_1 |\varphi^* >_2|e>_2
\end{equation}
and
\begin{equation}
|IV>=|\varphi >_1|g>_1 |\chi^* >_2|e>_2 \pm |\chi^* >_1|e>_1 |\varphi >_2|g>_2
\end{equation}
Each state leads to an absorption probability amplitude. These
amplitudes are indistinguishable because they are compatible with
the same absorption process and, consequently, they must be added to
get the full absorption probability. To add the amplitudes is
equivalent to consider a final state representing the superposition
of these states and to evaluate the transition rates with respect to
that final state, which reads
\begin{equation}
|\Phi _f>=N_f (a(|I>+|II>)+b(|III>+|IV>))
\end{equation}
Note that we must include the coefficients $a$ and $b$ in this
expression. If, instead, we would have taken the sum of the four
states with equal coefficients we would have erroneous results. For
instance, in the limit of no initial multi-particle superposition
($b=0$), the states $III$ and $IV$ would contribute to the matrix
element representing the transition from the initial to the final
state, a non-sense behavior. The normalization coefficient is
\begin{eqnarray}
N_f^{-2}=4(|a|^2+|b|^2)+4Re(a^*b<\psi^*|\varphi ^*><\phi |\chi>)+ \nonumber \\
4Re(ab^*<\chi^*|\phi ^*><\varphi |\psi>) \pm 4|a|^2Re(<\psi^*|\phi ^*><\phi
|\psi>)\pm \nonumber \\
4Re(a^*b<\psi^*|\chi ^*><\phi |\varphi>) \pm 4Re(ab^*<\varphi^*|\phi
^*><\chi |\psi>)\pm \nonumber \\ 4|b|^2Re(<\varphi^*|\chi ^*><\chi
|\varphi>)
\end{eqnarray}
Next, we evaluate the matrix element ${\cal M}= <0_{EM}|<\Phi _f
|\hat{U}|\Phi>|1_{EM}>$  representing the probability amplitude for
absorption. In this expression $\hat{U}$ is the evolution operator
of the complete system, and $1_{EM}$ and $0_{EM}$ represent the
states of the electromagnetic field of the light with $1$ and $0$
photons (remember that we have assumed a low intensity beam with
only one possible absorption). Because the two atoms do not interact
the evolution operator can be expressed in the factored form
$\hat{U}=\hat{U}_1 \otimes \hat{U}_2$. The atom-light interaction is
represented by the usual electric-dipole Hamiltonian. Moreover, at
the first order of perturbation theory the matrix element can be
expressed in the form $ <0_{EM}|<\Phi _f |\hat{H}_1 \otimes
\hat{H}_2|\Phi>|1_{EM}>$, with $\hat{H}_i$ the one-atom ($i=1,2$)
interaction electric-dipole Hamiltonian. It is proportional to ${\bf
D}.{\bf E}$, with ${\bf D}$ the total electric-dipole moment of atom
$i$ (as the atoms are identical both momenta are equal) and ${\bf
E}$ the electromagnetic field of the light.

Let us explicitly carry out the evaluation of one of the multiple
matrix elements included in ${\cal M}$, for instance,
\begin{eqnarray}
N_fN_0a <0_{EM}| _1<\psi^*|_1<e|_2<\phi|_2<g|  \hat{H}_1 \otimes \nonumber \\
\hat{H}_2  |\psi>_1|g>_1|\phi>_2|g>_2 |1_{EM}> =  N_fN_0a D<\psi^*|\psi>
\end{eqnarray}
with $D=D_0<e|{\bf D}|g>.<0_{EM}|{\bf E}|1_{EM}>$, where the
coefficient $D_0$ contains all the constant terms appearing in the
evaluation and whose explicit form is not relevant for our
discussion.

The rest of matrix elements can be calculated in the same way, and the full
matrix element becomes
\begin{eqnarray}
{\cal M}/2N_0N_fD= |a|^2(<\psi^*|\psi>+<\phi^*|\phi>)+ |b|^2
(<\varphi^*|\varphi> +<\chi^*|\chi>) + \nonumber \\
a^*b<\psi ^*|\varphi><\phi |\chi> + a^*b<\phi ^*|\chi><\psi |\varphi> +
ab^*<\varphi ^*|\psi><\chi |\phi> + \nonumber \\
ab^*<\chi ^*|\phi><\varphi |\psi>  \pm a^*b<\psi ^*|\chi><\phi |\varphi> \pm
ab^*<\varphi ^*|\phi><\chi |\psi> \pm  \nonumber \\
a^*b<\phi ^*|\varphi><\psi |\chi> \pm  ab^*<\chi ^*|\psi><\varphi |\phi> \pm
|a|^2<\psi ^*|\phi><\phi |\psi> \pm \nonumber \\ |a|^2<\phi ^*|\psi><\psi |\phi>
\pm |b|^2<\varphi ^*|\chi><\chi |\varphi> \pm |b|^2<\chi ^*|\varphi><\varphi
|\chi>
\end{eqnarray}
With this expression we can study the dependence of the absorption
rate, $|{\cal M}|^2$, on symmetrization and entanglement.

\section{Graphical representation}

In this section we represent the relative single absorption rate for
pairs of bosonic and fermionic atoms. It is defined as the
absorption rate normalized with respect to that of the same atoms in
a product state, $R=|{\cal M}|^2/ |{\cal M}_{pro}|^2$. The
absorption rate in a product state can be easily evaluated. If the
initial state is $|\eta>_1|g>_1|\mu>_2|g>_2$ the final one after
absorption is $(|\eta^*>_1|e>_1|\mu>_2|g>_2 + |\eta>_1|g>_1
|\mu^*>|e>_2)/\sqrt 2$. The matrix element reads ${\cal M}_{pro}
=D(<\eta^*|\eta>+<\mu^*|\mu>)/\sqrt 2$. We model the CM scalar
products with one recoil as $<\xi ^*|\Upsilon >=\alpha _0 <\xi
|\Upsilon>$ with $\alpha _0$ a constant coefficient. As the recoil
is in general a small effect we take $\alpha _0 =0.9$. Moreover, we
assume that it is independent of the CM state. With this
parametrization we have ${\cal M}_{pro} =\sqrt 2\alpha _0D$.

In addition to the parametrization for scalar products with a recoil
we must also specify these with two recoils. We take $<\xi ^*|\Upsilon ^*
>=\alpha ^2 <\xi |\Upsilon>$, where the coefficient has the form
$\alpha = \alpha _0 +(1-\alpha _0)c$. With this choice we guarantee
that if initially $|\xi>=|\Upsilon>$, at the end we will have
$<\xi^*|\Upsilon^*>=1$. If this condition would not be preserved we
would obtain erroneous results in the limit $|\xi> \rightarrow
|\Upsilon>$. In the particular case that $\xi$ and $\Upsilon$ do not
vary,$ <\xi |\Upsilon>$ is constant, we cannot take that limit and
we must use $\alpha = \alpha _0 +(1-\alpha _0) <\xi |\Upsilon>$. We
have extensively checked that this choice leads to the correct
results in the limits of complete overlapping, absence of
multi-particle superposition, ... Any other election gives errors in
some of these limiting situations.

\begin{figure}[H]
\centering
\begin{tabular}{cc}
\includegraphics[width=6cm,height=7cm]{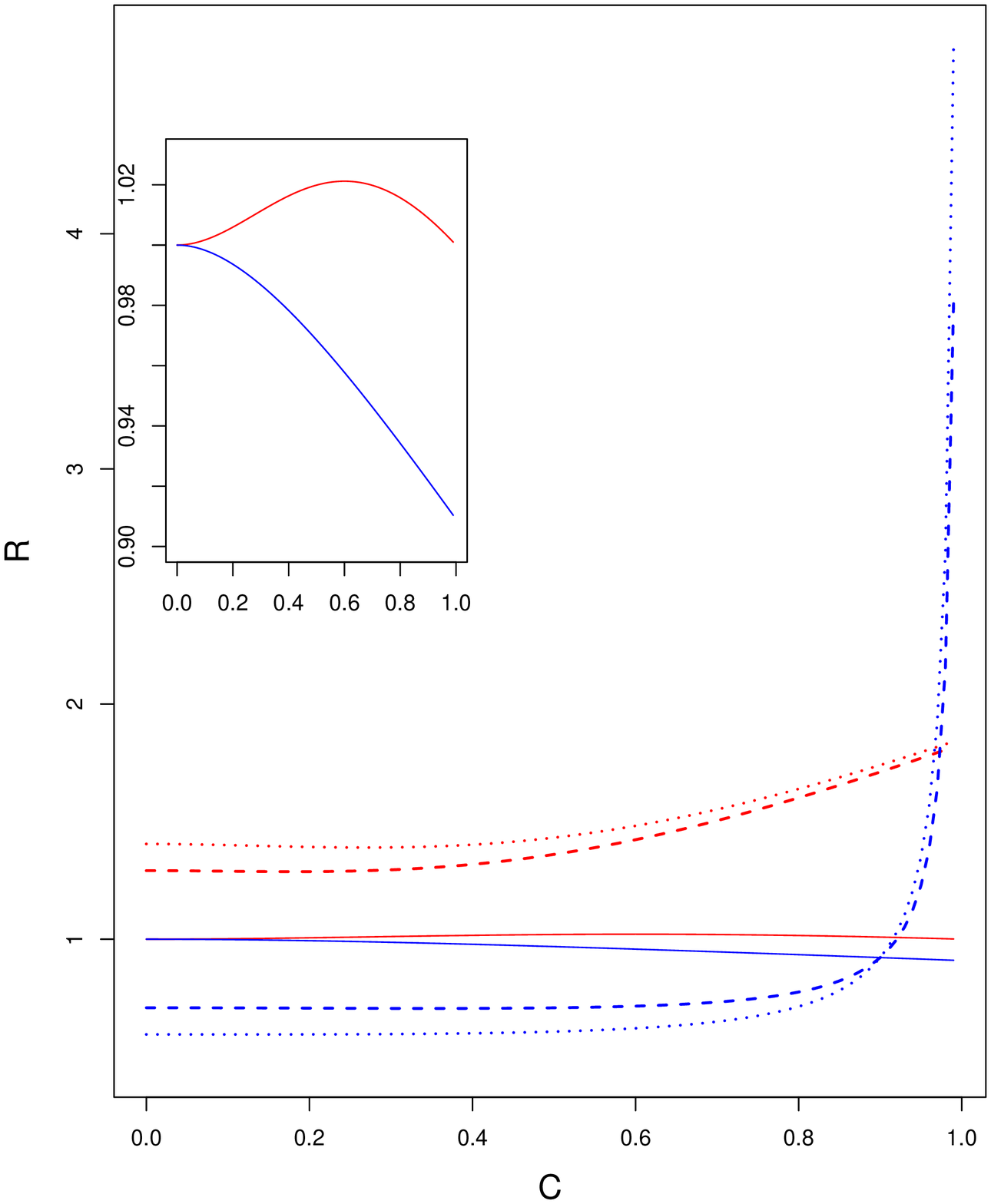}&
\includegraphics[width=6cm,height=7cm]{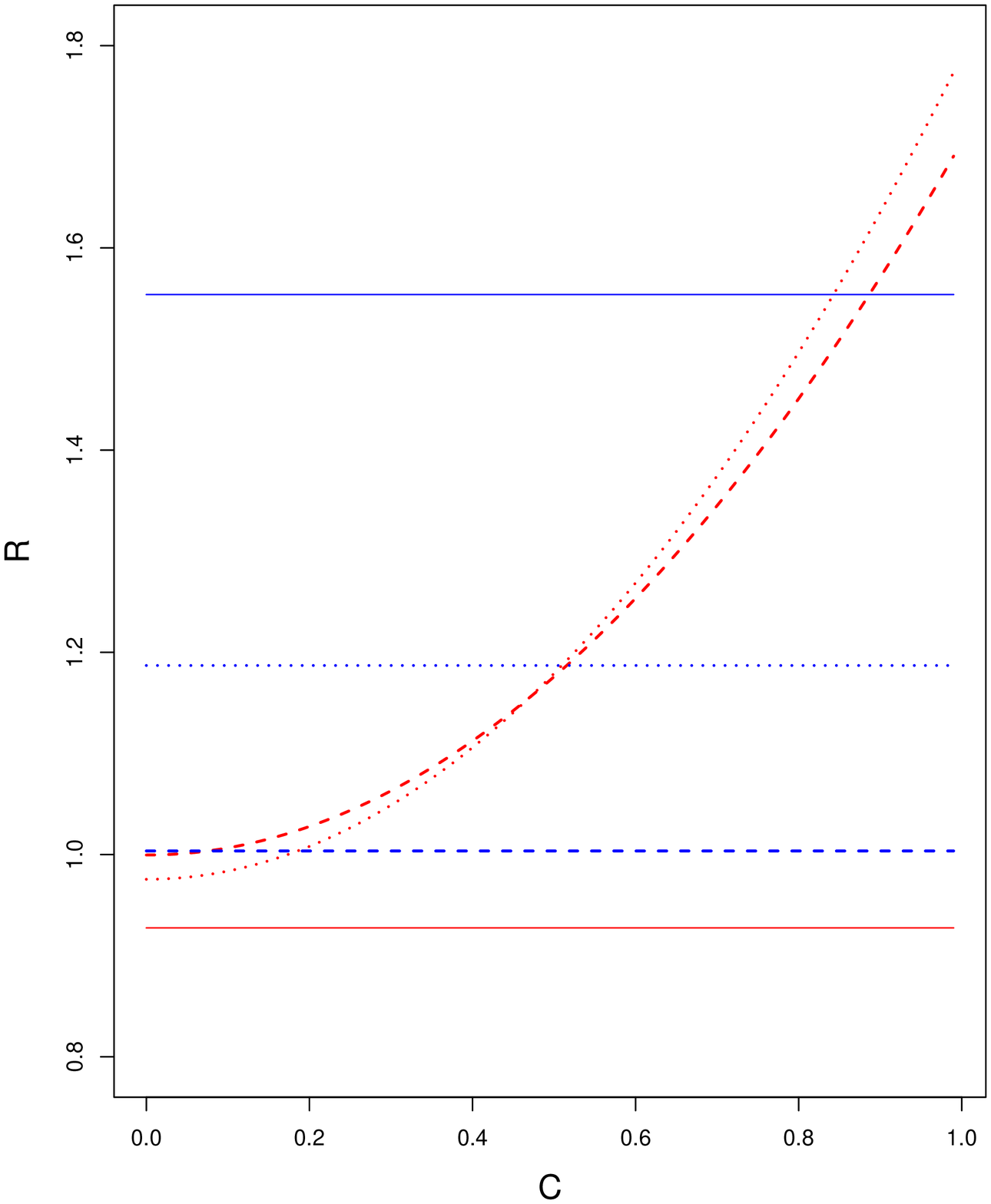}
\end{tabular}
\caption{Relative single absorption rate R versus initial overlap
$c$, both in arbitrary units, for the choices (i) left and (ii)
right. The red and blue curves correspond respectively to bosons and
fermions. The continuous, dashed and dotted lines represent the
cases $a=1, 0.8$ and $1/\sqrt 2$, with $|a|^2+|b|^2=1$. In order to
see more clearly the case with only symmetrization we include on the
left top corner of (i) the curves for this choice alone.}
\end{figure}

We represent the relative rate for four different choices of the CM
overlaps. For each choice we consider three different values of $a$
and $b$, , that is, of the relative weight of the two-atom states in
the superposition. One of the three cases is $b=0$ ($a=1$), which
corresponds to symmetrization without superposition. This way we can
compare the effects generated by symmetrization alone with those
present when symmetrization and superposition are simultaneously
taken into account. The four choices are: (i) $<\psi |\phi
>=c$, with $c$ real and varying in the interval $[0,1]$, $<\psi
|\varphi>=c$, $<\varphi|\chi>=0.9$,$<\psi
|\chi>=<\psi|\varphi><\varphi |\chi>$, $<\phi
|\varphi>=<\phi|\psi><\psi |\varphi>$ and $<\phi
|\chi>=<\phi|\varphi><\varphi |\chi>$. The last three relations are
equal for the rest of the cases. (ii) $<\psi |\phi >=0.8$,
$<\psi|\varphi>=c$, $<\varphi|\chi>=0.9$. (iii) $<\psi |\phi >=c$,
$<\psi|\varphi>=0.9$, $<\varphi|\chi>=c$. (iv) $<\psi |\phi >=0.8$,
$<\psi|\varphi>=0.9$, $<\varphi|\chi>=c$.

We represent choices (i) and (ii) in Fig. 2 and (iii) and (iv) in
Fig. 3. When only symmetrization is taken into account ($b=0$) we
have different behaviors for bosons and fermions in (i) and (iii)
(in (ii) and (iv) the overlap between $\psi$ and $\phi$ is constant
and there are not variations). For null overlap there are no
differences between them because the exchange effects disappear.
When the overlap is not null the situation changes. The relative
rate for fermions steadily decreases, progressively reaching larger
deviations with respect to the values associated with product
states. In contrast, the probability for bosons increases for small
overlaps reaching its maximum at intermediate values of $c$ and
decreasing later. For $c=1$ the initial bosonic state is
$|\psi>_1|\psi>_2$, a product one with no effects of symmetrization.
With the exception of $c=0$ and $c=1$ the absorption rates of
bosonic states are larger than those of product ones.

\begin{figure}[H]
\centering
\begin{tabular}{cc}
\includegraphics[width=6cm,height=7cm]{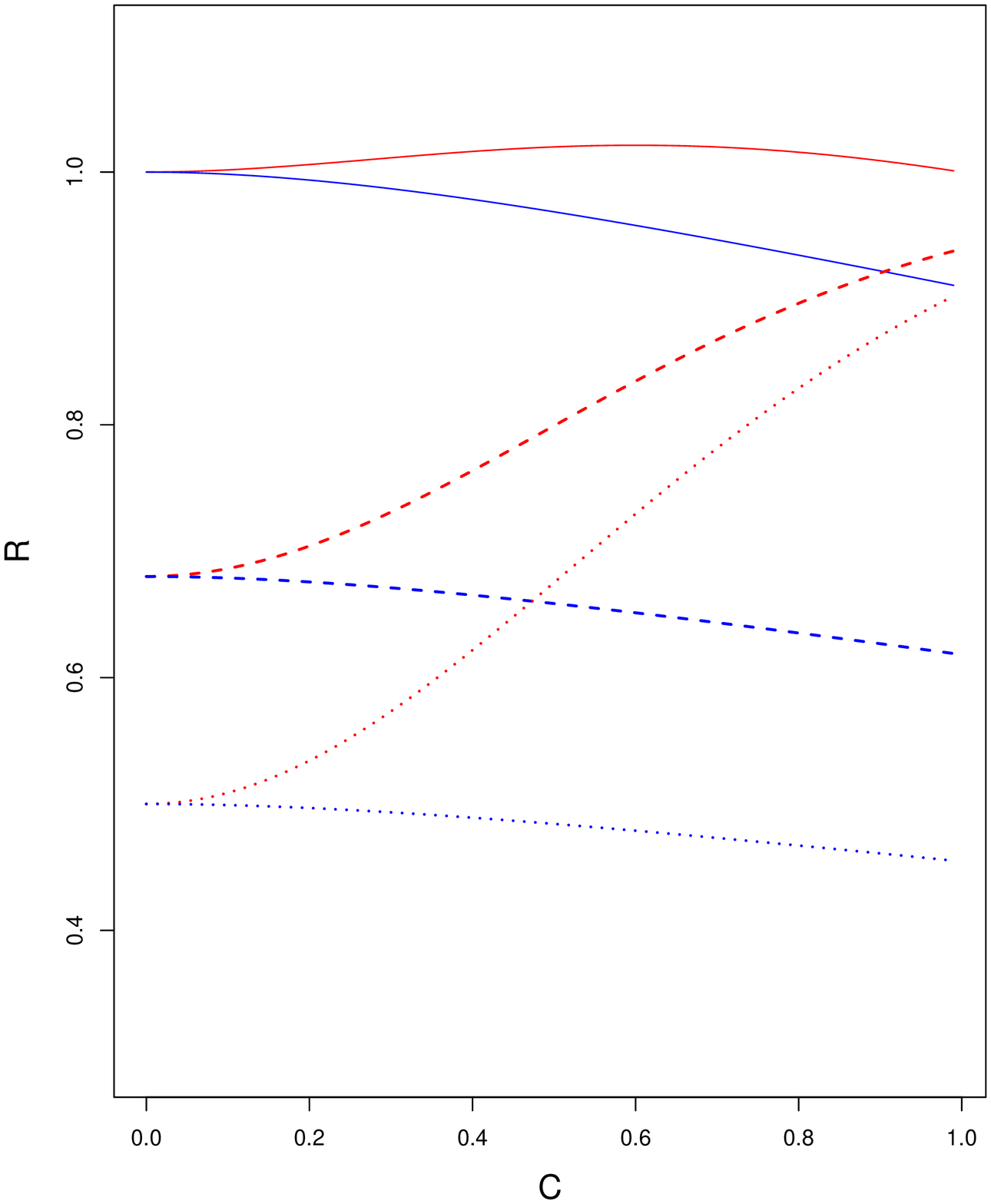}&
\includegraphics[width=6cm,height=7cm]{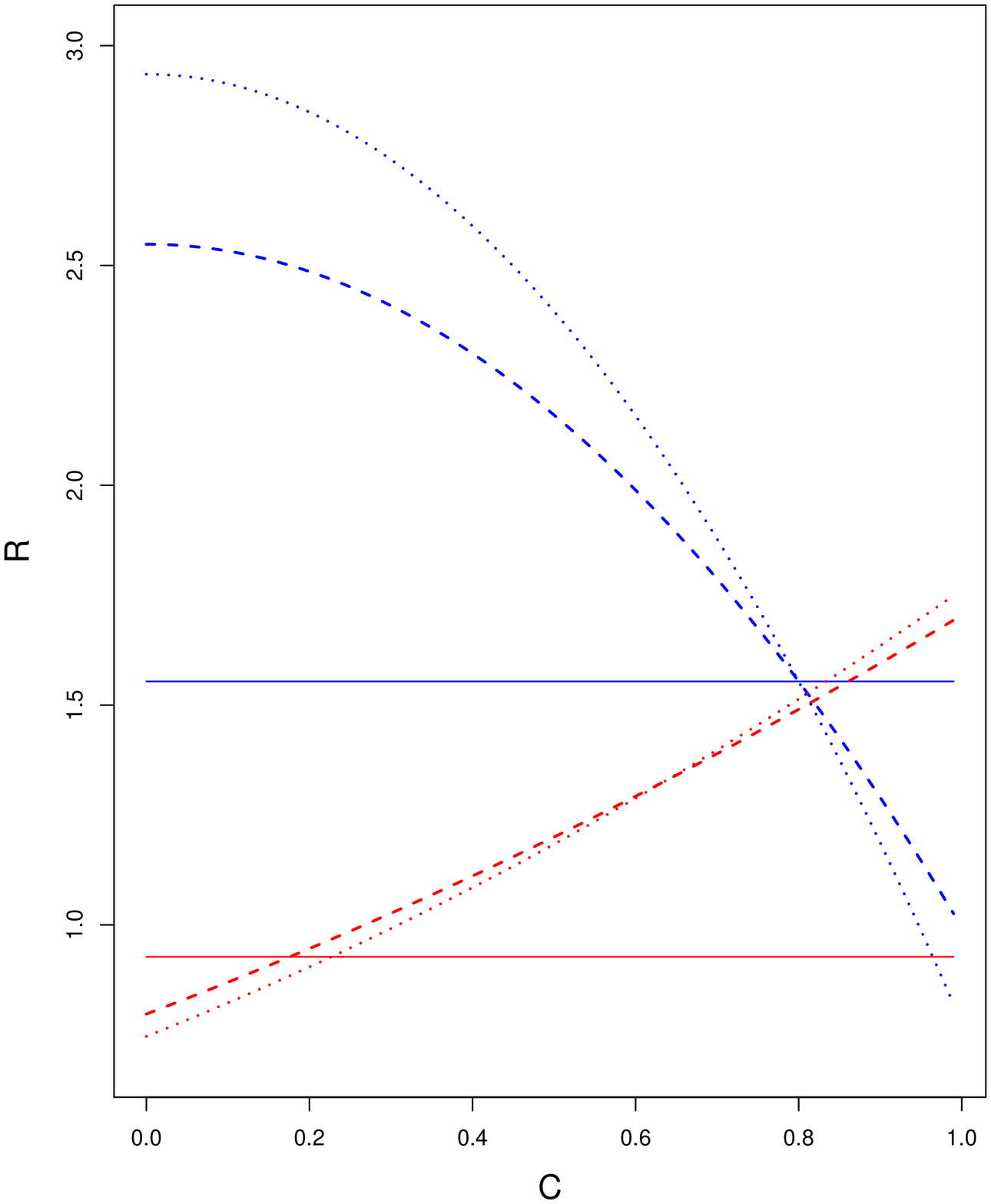}
\end{tabular}
\caption{The same as Fig. 2 for (iii) left and (iv) right. The
continuous, dashed and dotted lines correspond respectively to
$a=1,b=0$;$a=0.8,b=0.2$ and $a=b=0.5$ for (iii) and the same values
of $a$ with $|a|^2+|b|^2=1$ for (iv).}
\end{figure}

When multi-particle superposition is simultaneously taken into
account the general picture notoriously changes. First of all, note
that in various cases the relative rates of bosons and fermions for
$c=0$ are not equal. This is so because in these cases there is
overlap between some states of the superposition and the exchange
effects are present even in this limit.
For fermions in (ii) the relative rate does not depend on the
overlap, that is, the antisymmetrization effects have been canceled.
Multi-particle superposition can inhibit the antisymmetrization effects. This
situation corresponds to fixed overlap between the two states of
each term of the multi-particle superposition. In contrast, this
inhibition does not happen for bosons. Note that, although constant,
the values of the fermionic rates are very different depending on
the values of $a$ and $b$.

We have also observed (not represented here) the behavior that can
be considered complementary of the above inhibition. In the case
(iii) with $|a|^2+|b|^2=1$, for any choice of $a$ the fermion curves
coincide with that of $a=1$. There are antisymmetrization effects,
but the superposition ones have been canceled.
For both bosons and fermions the relative rates can be larger or
smaller than one. Another notorious difference between bosons and
fermions is that, in the presence of entanglement, in the first case
the relative rates always increase with the overlap. In contrast, in
the second case can increase (i), decrease (iii and iv) or be
constant (ii).

\section{Excluded states}

In addition to the examples presented in the previous section we
have analyzed many other situations. One specially interesting
occurs for the set of initial states $|\phi >=c|\psi>+d|\zeta>$,
$|\varphi >=e|\psi >+f|\zeta>$ and $|\chi >=g|\psi>+h|\zeta>$, where
$\zeta$ is a state orthogonal to $\psi$, $<\zeta |\psi>=0$. All the
coefficients obey the normalization conditions $|c|^2 +|d|^2=1$,
$|e|^2 +|f|^2=1$ and $|g|^2 +|h|^2=1$. Note that we do not include
the internal states of the atoms because they are irrelevant here.

We represent for fermions the particular case $e=f=1/\sqrt 2$ and
$|\chi>=c|\varphi>+ d|\varphi ^{\perp}>$, with $\varphi ^{\perp}$
orthogonal to $\varphi$. If we take $|\varphi ^{\perp}>=
(|\psi>-|\zeta>)/\sqrt 2$, the coefficients are $g=(c+d)/\sqrt 2$
and $h=(c-d)/\sqrt 2$. The results are presented in Fig. 4. For
superpositions with equal weights we observe a non-regular behavior.
This behavior can be explained analytically invoking the
unnormalized form of Eq. (\ref{eq:uno}) for our set of states:
\begin{equation}
|\bar{\Phi}>=(ad+b(eh-fg))(|\psi>_1|\zeta>_2-|\zeta>_1|\psi>_2)
\label{eq:exc}
\end{equation}
Replacing the values used in the graphical representation we see
that for $a=b$, $|\bar{\Phi}>=0$. For the normalized state we have
$|\bar{\Phi}>/\parallel  |\bar{\Phi}> \parallel =0/0$, an undefined
form that is typical from excluded states.

This is the reason for the observed non-regular behavior. When we
introduce this state in the graphical representation program, it
deals with an undefined expression of the type $0/0$. However,
because of numerical round-off errors can give an output (the pieces
of line that are represented), which anyway are null. In
contrast, for other values the programm gets an undefined answer and
cannot represent a value.

This is not the only state with this form. From Eq. (\ref{eq:exc})
we can see that any state whose coefficients obey the relation
$ad+beh-bfg=0$ has the same undefined form. As a particular case, we
have that when there is not superposition, $b=0$, the above
condition reads $d=0$, or equivalently, $\phi =\psi$. For
non-entangled states this is the usual Pauli's condition for
exclusion. We have found a set of excluded entangled states. The
exclusion condition in this case is different from the usual one and
does not require the one-particle states to be equal. The Pauli-type
condition is only recovered for non-entangled states.

\begin{figure}[H]
\center
\includegraphics[width=10cm,height=5cm]{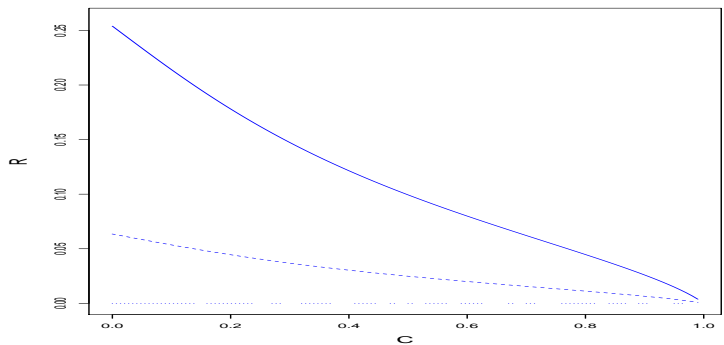}
\caption{The same as in Fig. 2 for fermions in the states discussed
in this section. The continuous, dashed and dotted lines correspond
to $a=0.64,0.67$ and $1/\sqrt 2$, with $|a|^2+|b|^2=1$.}
\end{figure}

The possibility of excluded entangled states beyond the scope of
Pauli's principle \cite{pau} has been previously considered in more
theoretical grounds \cite{yar}. The fundamental idea in that
proposal is that in entangled systems you can define the
multi-fermion state but not the one-fermion ones and, consequently,
the exclusion conditions must be based on different considerations.
The results of this paper show that there are non-standard excluded
entangled states in realistic systems.

\section{Conclusions}

The interplay between symmetrization and entanglement is a subtle
and difficult subject. Topics like the definition of entanglement
measures in systems of identical particles or the conversion of the
identity correlations into useful entanglement have deserved a lot
of attention, but there are other problems, like the physical
modifications associated with their joint effects, that remain
almost unexplored.

We have shown that there is not an universal behavior of the
absorption rates when symmetrization and multi-particle
superposition act at once. Depending on the parameters of the
problem, the coefficient $a$ and $b$ in the superposition and the choice
of overlaps between the spatial CM states, we have a big variety of
absorption rate patterns. The presence of two-particle
superpositions drastically changes the analytical form of the
symmetrized absorption rates of  identical particles. The observed
effects cannot be considered as the addition of those associated
with the contributions of symmetrization and superposition. One
could associate the exchange terms (those with the sign $\pm$) with
the effects of symmetrization and the rest with superposition.
However, the normalization coefficients contain exchange- and
non-exchange-type contributions and the above separation is not
possible. We must also remark the notorious differences between the
behavior of bosons and fermions in all the cases.

We have also found excluded entangled states beyond the scope of
Pauli's principle. From the experimental point of view, these states
could be tested noting that fixing the values of the overlaps and
approaching the limit $a=b$ we would see a progressive decreasing of
the absorption rates towards a null value (see Fig. 4). The
arrangement discussed in the second section seems to be well suited
to implement the experimental scheme.

\end{document}